\documentclass[manuscript,format=acmsmall, screen=true,review=false]{acmart}

\AtBeginDocument{%
  \providecommand\BibTeX{{%
    \normalfont B\kern-0.5em{\scshape i\kern-0.25em b}\kern-0.8em\TeX}}}

\setcopyright{acmcopyright}
\copyrightyear{2018}
\acmYear{2018}
\acmDOI{10.1145/1122445.1122456}

\acmConference[]{}{}{}
\acmBooktitle{}
\acmPrice{}
\acmISBN{}

\usepackage{graphicx}
\begin{document}

\title{Rihanna versus Bollywood: Twitter Influencers and the Indian Farmers’ Protest}

\author{Dibyendu Mishra}
\email{v-dibmis@microsoft.com}
\author{Syeda Zainab Akbar}
\email{v-syakba@microsoft.com}
\author{Arshia Arya}
\email{t-aryaarshia@microsoft.com}
\author{Saloni Dash}
\email{t-sadash@microsoft.com}
\author{Rynaa Grover}
\email{t-rgrover@microsoft.com}
\author{Joyojeet Pal}
\email{joyojeet.pal@microsoft.com}
\affiliation{%
  \institution{Microsoft Research}
  \country{India}
}

\begin{abstract}
A tweet from popular entertainer and businesswoman, Rihanna, bringing attention to farmers' protests around Delhi set off heightened activity on Indian social media. An immediate consequence was the weighing in by Indian politicians, entertainers, media and other influencers on the issue. In this paper, we use data from Twitter and an archive of debunked misinformation stories to understand some of the patterns around influencer engagement with a political issue. We found that more followed influencers were less likely to come out in support of the tweet. We also find that the later engagement of major influencers on the side of the government's position shows suggestion's of collusion. Irrespective of their position on the issue,  influencers who engaged saw a significant rise in their following after their tweets. While a number of tweets thanked Rihanna for raising awareness on the issue, she was systematically trolled on the grounds of her gender, race, nationality and religion. Finally, we observed how misinformation existing prior to the tweet set up the grounds for alternative narratives that emerged.   
\end{abstract}



\maketitle
\section{Introduction}
On February 2, 2021, American musician Rihanna posted a tweet, asking “why aren’t we talking about this?!". \#FarmersProtest” refers to the ongoing protests against new laws in India that change the processes of agricultural sale and storage. The tweet immediately went viral, garnering over half a million engagements by the end of the day. It also triggered messaging from other internationally known figures including climate activist Greta Thunberg, lawyer Meena Harris, and media personality Mia Khalifa.

The following day, a series of coordinated messages appeared from several Indian entertainers and sportspersons, calling the Twitter activity of the previous day propaganda. The event has triggered much discussion on the perceived influence that the Indian government has on celebrities, and in turn the perceived influence of celebrities on political opinion. In this short paper, we analyze the key trends that occurred in the two days following the social media posts, and consider the impact these have had on the social media following of the individuals.

Our key findings are as follows. First, we find that the more followed a celebrity, the less likely they engaged in support of the tweet by global women influencers who spoke on the farmer protests. Second, we find strong suggestion of collusive tweeting from the key Indian celebrities, based on the timestamps of their tweeting, and the contents of the text. Third, we find evidence of systematic trolling of the key figures, particularly Rihanna, and that this trolling is most common on issues of gender, ethnicity/race, and religion. Fourth, we find that speaking on the issue, on all sides of the political spectrum, resulted in increased following for the respective Twitter account holders’, even when they were subject of high degree of trolling. Finally, we find that the misinformation related to the Farmer protests that existed prior to the most recent round of Twitter activity set up the grounds for three major narratives that have appeared: (1) that there are close parallels to the way misinformation is framed in the farmers’ protest case with other situations with anti-government protests or speech (CAA/NRC, COVID-19), (2) that those who are in favour of the farmer protests are anti-national and/or driven by foreign influence, using Muslim and Khalistani as slurs, and (3) around a series of false images or connections between figures from Indian politics and key members of the Adani and Ambani families.

\section{Data and Methodology}

\subsection{Data}
Using Twitter's search API we collected 187,045 tweets around two key entities: 1) Rihanna's tweet on the farmer's protests on February 2, 2021 2) The hashtags \#IndiaAgainstPropaganda and \#IndiaTogether. For 1 we collected all quote tweets and comments in the 12 hour interval after the time of authoring i.e February 2, 9 PM. Since we wanted to focus on public and influential figures on social media, for 2 we collected all tweets from verified handles and handles from a publicly available list of Indian politicians\cite{nivaduck} and a list of their influential friends with the said hashtags on February 3. 

While there may be cases where one would mask the IDs of private persons, we take the position in this paper that all of the research here is on individuals who are public figures such as journalists, politicians, and celebrities posting publicly, thus we use aggregations of their impact.

\textit{Twitter Trending Hashtags and Topics}: We collected the top 50 trending hashtags and topics for each hour from February 2, 2021 to February 4, 2021. Along with timestamps, we also collected the volume of tweets corresponding to each topic in a given hour. These were annotated for their relevance to the Farmer's protests by a group of 2 researchers and had a inter coder reliability score of 0.876. The hashtags disagreed upon were reconciled and re-annotated later. 

\textit{Twitter Followers Count}: We collected profile information on each day from 31st Jan to 4th of Jan for every twitter handle from Nivaduck's publicly annotated list and a list of their influential friends. 

\textit{Misinformation}: For debunked misinformation, we borrow our data for the period of September 15th, 2021 to February 4, 2021 from an archive maintained by Tattle Civic Technology (under ODBL license) \cite{Tattle} for the IFCN certified fact-checker Boomlive. We manually coded these stories for Farmers' protest related fact-check stories and found 108 unique stories directly related to the case. However, as our sample is limited to stories that have been debunked we are restricted by the limited resources and selection criteria of fact-checking organisations. 
\subsection{Stance Classification}
\begin{table}[!htb]
\small
\begin{tabular}{c|c}
\hline
Stance & Keywords \\ \hline
Pro & \begin{tabular}[c]{@{}c@{}}supporting, thankyou, thanks, recognition, forefront, limelight, visibility,\\ siding, injustice, awareness,  fascist, dictator, atrocities\end{tabular} \\ \hline
Anti & \begin{tabular}[c]{@{}c@{}c@{}}affairs, sensitive, involve, interfere, hijacked, leftist, dumb, troll, terrorism, \\ business, hefty, dolts, staged, Chinese, poke, idiots, interrupt, meddle, nose, \\ antiindia,  interference, poking, antinational, anarchists\end{tabular} \\ \hline
\end{tabular}
\caption{\label{tab:Table 1}\textbf{Keywords for each stance}}
\end{table}
To classify tweets into stances, we define an initial set of high precision keywords and expand on them using Word2Vec\cite{mikolov2013efficient}. For each stance, i.e either supporting the farmer's protests/Rihanna or opposing them, we create a list of keywords that indicate either stance. Next, we filter all tweets containing any of these keywords and train a Word2Vec model on them. Using a cosine similarity based criteria\cite{vijayaraghavan2016automatic}, we expand on the seed set of words for each stance with the closest words to their vector representations. Finally, we used the list of keywords mentioned in Table \ref{tab:Table 1} to match tweets for stance relevance.

\subsection{Hate Speech Detection}
To detect and analyse the hate speech against Rihanna, we used a lexical filtering approach. We collected English and Hindi hate terms from an online respository called HateBase \cite{HateBase}. Along with the words and their meanings, the API mentions the grounds on which each word is offensive. Using these dictionaries of Hindi and English hate words, we looked for their occurrences in the replies and quotes of Rihanna’s tweet. To contextualise the repository to this particular case we carefully curated the terms from the API and removed words like ‘queen’, ‘girl’ and ‘Punjab’ which appeared in a more positive light.  

\section{Celebrity Engagement and Collusion}

We collected (using Twitter’s search API) about 100k responses to Rihanna’s tweet by means of comments and quote tweets in the first 12 hours after she authored it. In Figure 1 we see that the overwhelming engagements of public figures in India were supportive of Rihanna, but the vast majority of people who tweeted in support of Rihanna though influential, were not among the most influential Indian celebrities. Only one Indian celebrity with over 1 million Twitter followers sent an immediate endorsement (quoted or replied), actor Swara Bhaskar, while Gul Panag sent a neutral toned message.

\begin{figure}[!htb]
  \centering
  \includegraphics[width=0.7\linewidth]{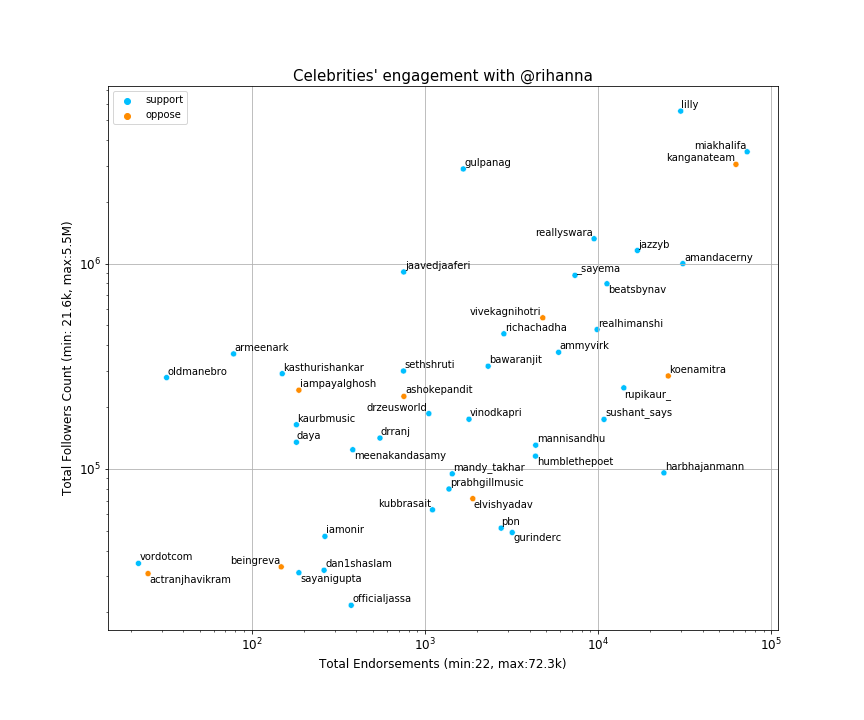}
  \caption{ \textbf{Celebrities’ Engagement with Rihanna’s Tweet}: Celebrities who tweeted in support (blue) and opposition (orange) of Rihanna are visualised as a function of the total endorsements (retweets + likes) on their tweet vs their total follower count.}
\end{figure}

The more important trend came from a number of Indian-origin public figures from various other parts of the world, particularly British and North American entertainers of Punjabi origin including Lilly Singh, Jazzy B, Ammy Virk, Nav, Rupi Kaur, Panjabi by Nature, Gurinder Chaddha etc. This in part aligns with the support for the farmers’ cause within that constituency, which has also been successful in getting political representatives from areas with significant Punjabi populations including Manchester (Yasmin Qureshi), Ontario (Gurratan Singh), and the California central valley (US Rep Jim Costa) tweeting about the Farmer protests.

The Indian film industry was mainly represented by a smaller number of public facing figures such as actors or musicians who have all embraced their political positions publicly (Jaaved Jaaferi, RJ Sayema, Richa Chaddha etc.), and a larger proportion of people behind the scenes such as filmmakers or producers (Vinod Kapri, Onir, Guneet Monga etc.). On the pro-government end of the spectrum, filmmakers Vivek Agnihotri and Ashok Pandit, and actors Payal Ghosh and Koena Mitra engaged with Rihanna within the first 24 hours of the posting. Overall, we find that while a range of entertainers and journalists got engaged in the activity surrounding the tweets, almost no sportspersons got involved in support of Rihanna, except for badminton champion Jwala Gutta who posted an emoji in response. The second category of missing influencers is people in the business world. While industrialists are frequently vocal on Twitter on issues related to the nation, they eschewed involvement in this case. The only sportsperson of significant following who tweeted against Rihanna, prior to the next day’s events, was former cricketer Pragyan Ojha.

\begin{figure}[!htb]
  \centering
  \includegraphics[width=0.8\linewidth]{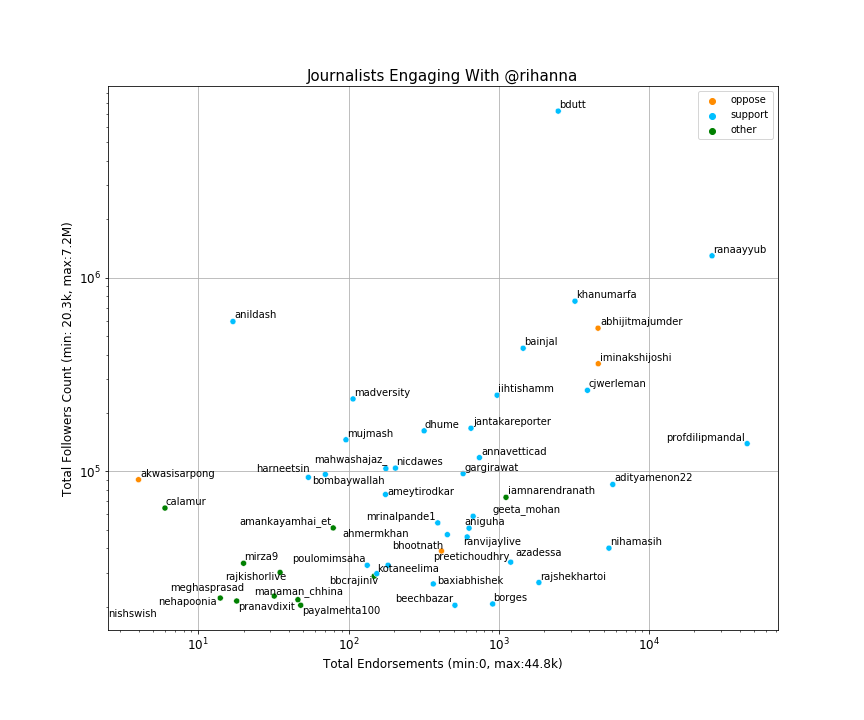}
  \caption{ \textbf{Journalists’ Engagement with Rihanna’s Tweet}: Journalists who tweeted in support (blue) and opposition (orange) of Rihanna are visualised as a function of the total endorsements (retweet + likes) vs their total follower count.}
\end{figure}
A look at the journalists or public commentators who engaged with Rihanna in Figure 2 shows that the majority were supportive of her position, with a few exceptions such as Abhijit Majumder and Meenakshi Joshi. Dilip Mandal was the journalist/commentator who had the maximum amount of engagement related to Rihanna, though he was also the individual who tweeted about Rihanna the most from among the key figures in mainstream media. Rana Ayyub had the highest engagement from a single tweet.

\section{Coordinated Tweeting}

On February 3, when the bulk of Indian celebrities got involved, we see a significant spread of highly followed persons chiming in, several with the same few hashtags such as \#IndiaAgainstPropaganda and \#IndiaTogether. In Figure 3 we plotted the key celebrities who tweeted with the volume of their online following on one axis and the extent to which their tweets were endorsed. To calculate the total endorsement, we used aggregate engagement, thus the net engagement of all retweets, quote tweets, and replies of all tweets on the issue. We also plotted the engagement by Mia Khalifa and Rihanna to compare their impact against that of the Indian celebrities. We find that Rihanna has a massively larger impact (almost 3x) of any of the Indian celebrities in terms of net engagement, and that Mia Khalifa is second only to Sachin Tendulkar despite less than a tenth of his online following. Kangana Ranaut is a close third, though she tweeted more often on the issue than any other person from among the entertainers.
\begin{figure}[!htb]
  \centering
  \includegraphics[width=0.8\linewidth]{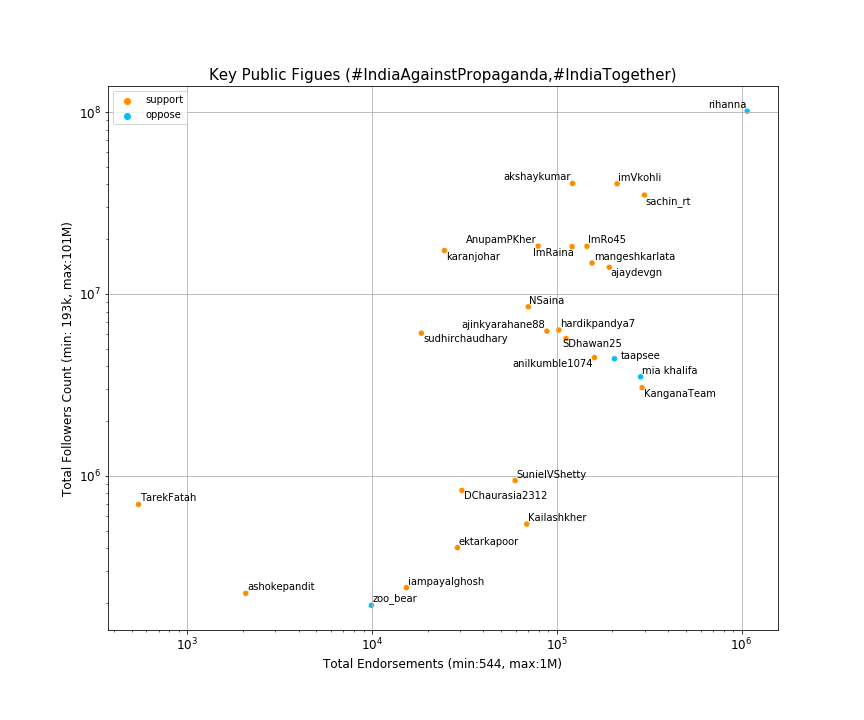}
  \caption{\textbf{Key Public Figures Tweeting \#IndiaTogether or \#IndiaAgainstPropaganda}. Those who tweeted in support (blue) and opposition (orange) are visualised as a function of the total endorsements (retweet + likes) vs their total follower count.}
\end{figure}

\begin{figure}[!htb]
  \centering
  \includegraphics[width=0.8\linewidth]{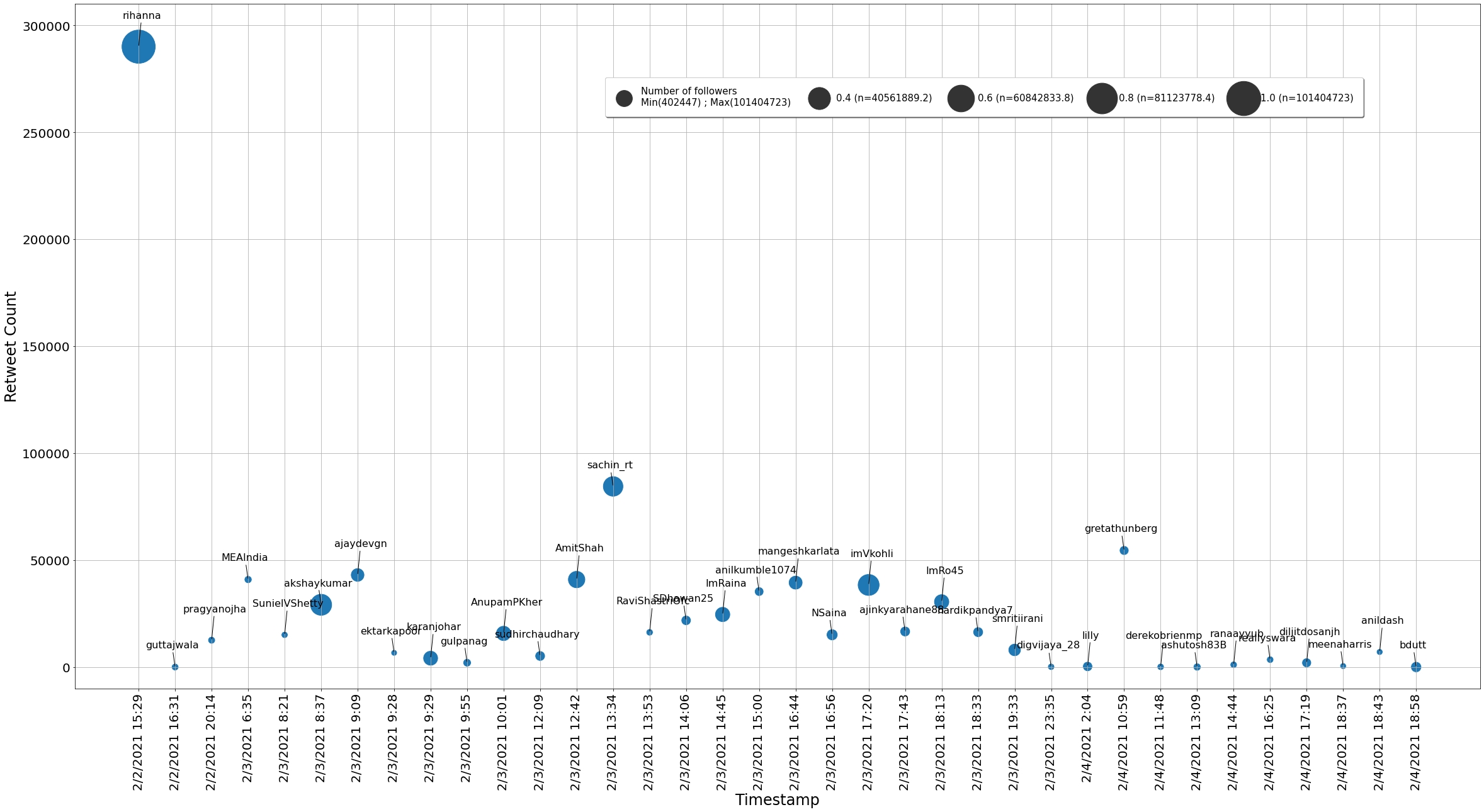}
  \caption{\textbf{Timeline of tweets by key influencers}}
\end{figure}

There has been much discussion on the likelihood of collusion in the social media posting of the key figures. We also see some important patterns in how the tweeting from the celebrities in India went out, once we plot the tweets on a timeline in Figure 4, we see a pattern of film stars first starting with a tweet from the Ministry of External Affairs at 6.35 AM. Starting at around 8 AM, Sunil Shetty was the first to tweet, followed by Akshay Kumar, Ajay Devgun, Ekta Kapoor, Karan Johar and Anupam Kher at roughly 20 min intervals. A second round began with a tweet from Amit Shah at 12.42 PM, which was followed by Sachin Tendulkar the next hour, with several cricketers following at roughly a 20 min interval again including Ravi Shastri, Shikhar Dhawan, Suresh Raina and Anil Kumble. The third wave began with singer Lata Mangeshkar at 4.44 PM, followed by badminton player Sania Nehwal, and cricketers Virat Kohli, Ajinkya Rahane, Rohit Sharma, ending with Hardik Pandya at 6.33 PM. So, we see that each wave had at least one massively popular social media star with 35 Million or more followers (Akshay Kumar, Sachin Tendulkar, Virat Kohli).

\begin{figure}[!htb]
  \centering
  \includegraphics[width=0.8\linewidth]{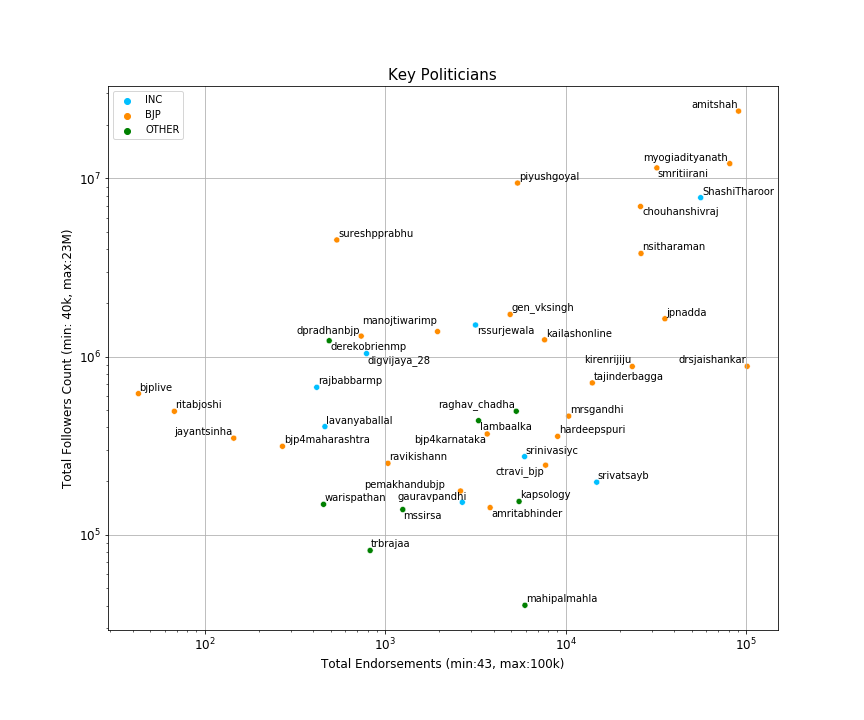}
  \caption{\textbf{Key Politicians Engaging with Farmers’ Protests Issue After Rihanna’s Tweet}: Politicians who belong to INC (blue), BJP (orange)  and other parties (green) are visualised as a function of the total endorsements (retweet + likes) vs their total follower count.}
\end{figure}
In Figure 5 we also see that various politicians tweeted about the issue, and while the overall spread of politicians who tweeted about it (Figure 6) is dominated by non-BJP politicians, the politicians from the opposition parties were mainly less popular leaders with limited following, whereas in the BJP, several top brass leaders tweeted about the subject. MEA S Jaishankar was the most engaged overall, while Amit Shah and Yogi Adityanath, JP Nadda, and Nirmala Sitharaman all were significantly engaged. On the opposition side Shashi Tharoor, with a carefully worded message was the most engaged, whereas Akhilesh Yadav was among the earliest to retweet Rihanna. Congress leader Rahul Gandhi and Prime Minister Narendra Modi were both conspicuous in their absence.

\begin{figure}[!htb]
  \centering
  \includegraphics[width=0.8\linewidth]{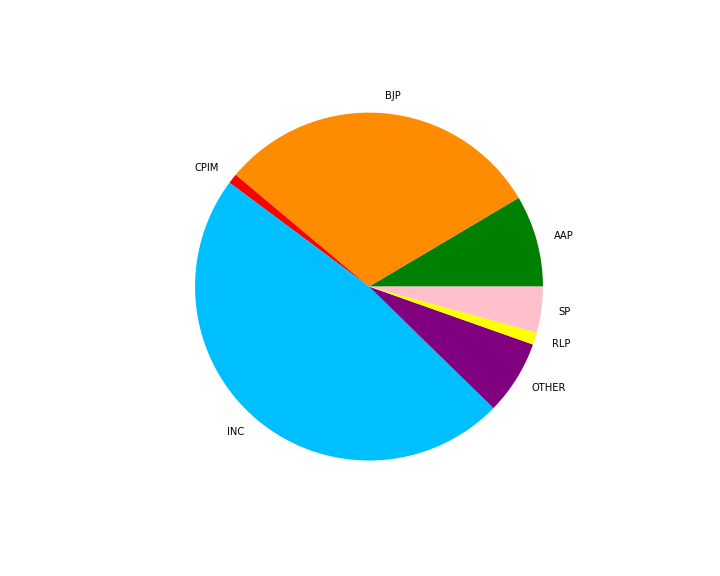}
  \caption{\textbf{Distribution of Politicians Tweeting About Farmers’ Protest After Rihanna’s Tweet}}
\end{figure}

\section{Trending Topics and Hashtags}

A look at the timeline of trending topics(Figure 7) and hashtags shows that the first 24 hours following Rihanna’s tweet, there was much momentum on the side batting for the protesting farmers against the government. However, that the next day, following the of the Indian celebrities, the tide turned towards the side opposing Rihanna’s position.
\begin{figure}[!htb]
  \centering
  \includegraphics[width=\linewidth]{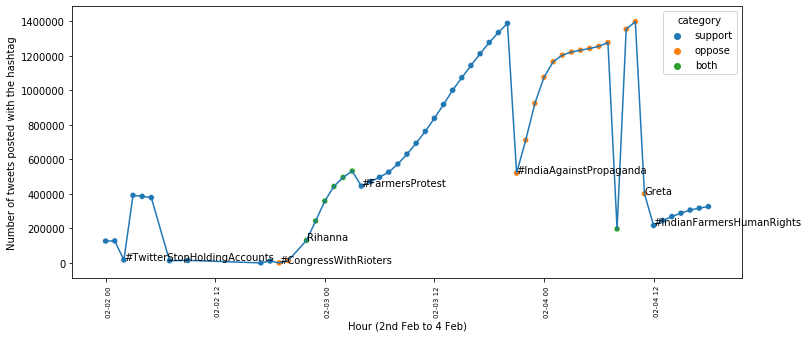}
  \caption{ \textbf{An Hourly Timeline of Top Trending Hashtags}: Starting from 2nd Feb up to 4th of Feb. The x axis represents the number of tweets posted with the top trending hashtags in that hour. The colors of the markers depict whether majority of the tweets with the hashtags were in support, oppositon or both with respect to the Farmers’ Protests.}
\end{figure}
A look at the trending hashtags reinforces this. In Figure 8, we see that the overall trending hashtags are much more on the side of the government than against the government. These are not necessarily a full spread of hashtags, but those that trended during the period of the events. For a hashtag to trend, there needs to either be a very organic push, or concerted astroturfing. Thus, while Rihanna was a trending topic much of the time, there was only one trending hashtag around her, whereas Greta Thunberg was systematically targeted through multiple hashtags along with Mia Khalifa. While the generic \#FarmersProtest, a hashtag that has trended before the events of this week, specifically constructed for this – \#IndiaTogether, \#IndiaStandsTogether and \#IndiaAgainstPropaganda all trended. This speaks to the effectiveness of the planning and branding of these in the short reaction time available.
\begin{figure}[!htb]
  \centering
  \includegraphics[width=0.5\linewidth]{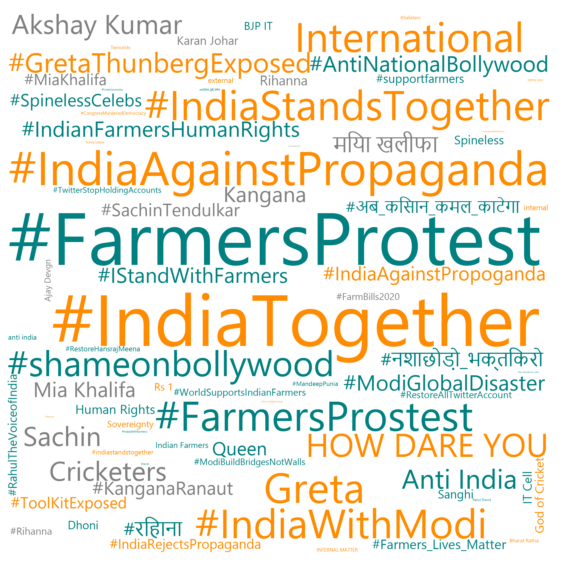}
  \caption{ \textbf{Overall WordClouds of Trending Hashtags}: The prominence of a hashtag/topic is scored based on the number of tweets and the number of hours it trended(Teal: Support Farmers’ Protests, Orange: Oppose Farmer Protests, Grey: Both)}
\end{figure}

\section{Targeting of Rihanna}

In Figure 9 the wordclouds highlight the discourse on both sides of the argument. The pro side focuses on words around thanking, support, and awareness, since their goal is to bring wider acceptance of the cause, whereas the oppositional discourse focuses much more strongly on a discourse of ownership of the issue. Thus, it uses terms such as business (i.e. mind your own business), interfere, nose (i.e. poke nose) or terms intending to delegitimize the opposition such as Khalistan, paid, fake etc.

\begin{figure}[!htb]
  \centering
  \includegraphics[width=\linewidth]{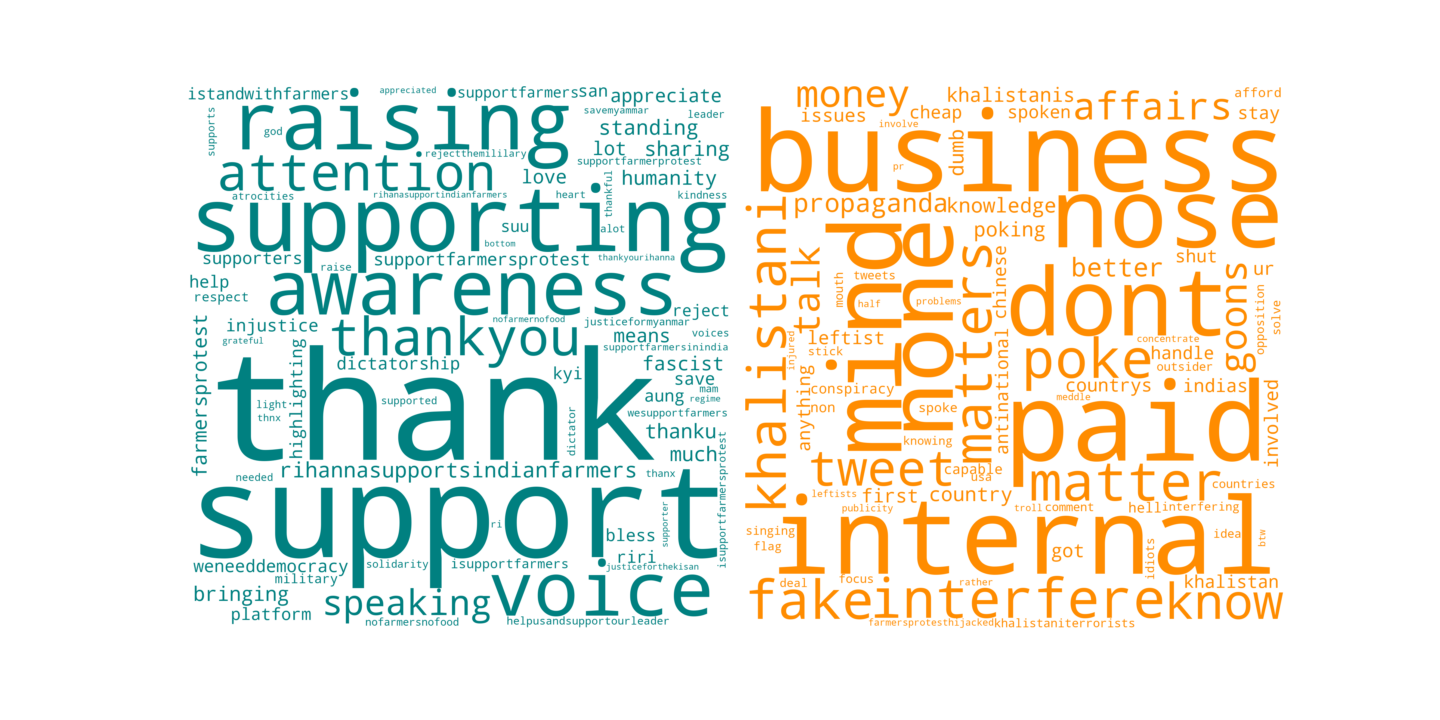}
  \caption{\textbf{Wordclouds of Opposite Stances for Rihanna’s Tweet}: Teal depicts supporting tweets whereas orange depicts opposition.}
\end{figure}

\begin{figure}[!htb]
  \centering
  \includegraphics[width=0.8\linewidth]{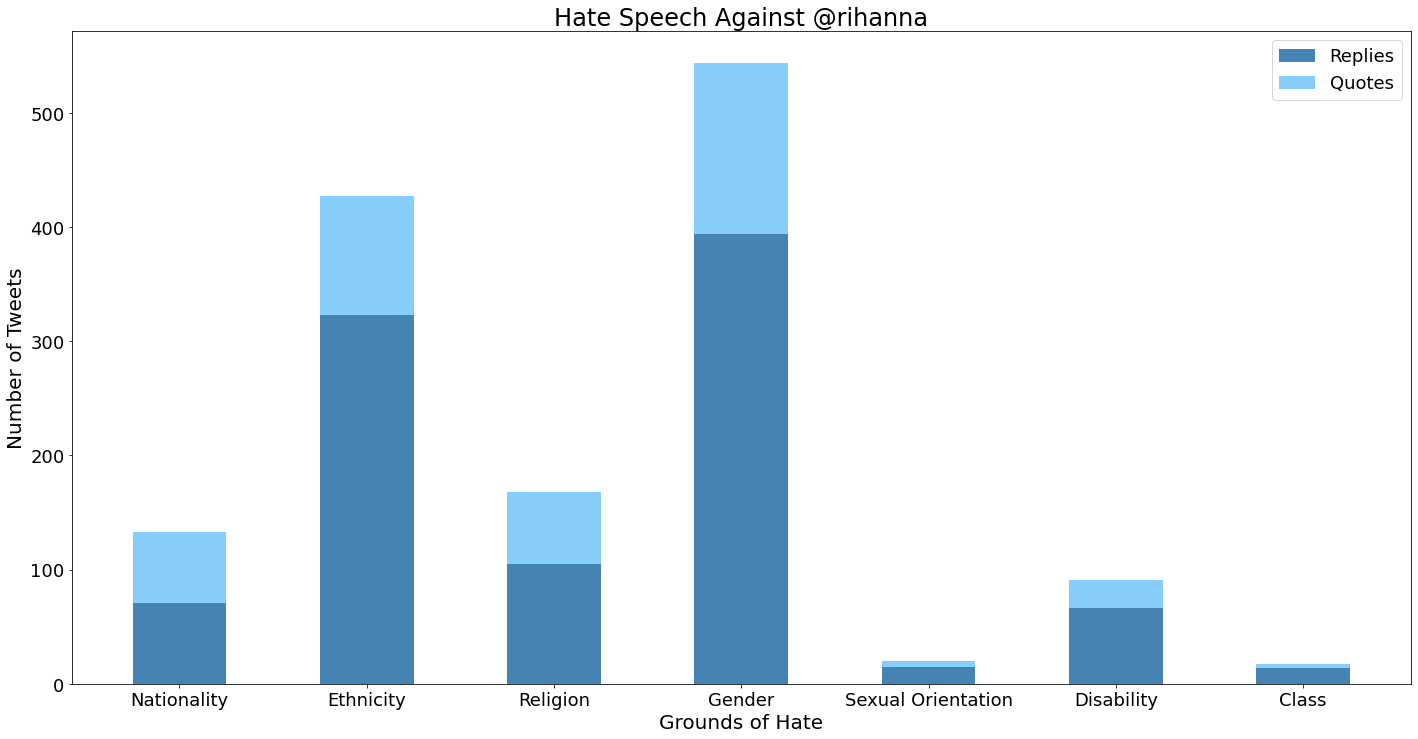}
  \caption{\textbf{Hate Speech Targeted at Rihanna}: Classification of comments and quotes on Rihanna’s tweet detected to have hate speech in them}
\end{figure}
The patterns seen in the data in Figure 10 show that the main line of attack against Rihanna was gendered, attempting to delegitimize her as a woman, and second, for her ethnicity, in this case primarily race. As much news reporting has already shown, there has been significant use of racist language against Rihanna following her tweet. While the nationality of the key persons involved in the issue (Rihanna, Greta Thunberg, Amanda Cerny, Meena Harris) has been raised significantly in the “interference from outsiders” arguments put forth, there is little doubt that the key opponents being women is a central part of the counterattack from those opposed.

\section{All Publicity is Good Publicity}
To check the impact of engagement with the issue, we decided to see whose followers increased or decreased following the events between Feb 2 and 4 in Table \ref{tab:Table 2}. The results are interesting. While our results do not conclusively prove causality, we see that across both sides of the issue, there is an increase in following of the accounts irrespective of their stance. We selected a subset of the accounts that received a lot of attention during the engagement around the issue. From this subset, the highest increase in following have been for the Ministry of External Affairs (@meaindia) and Mia Khalifa (@miakhalifa), both of which had in excess of an 1581 \%  (more than 15 times what they typically add in two days) increase over their average daily increase in following for the period just preceding the incident. Even for accounts like that of Sachin Tendulkar (@sachin\_rt) who was widely trolled, the increase is more than 2x of the daily increase in following from the previous period. Television Anchor Sudhir Chaudhary (@sudhirchaudhary) who came down heavily in favour of the official position had an increase of almost 4x of his typical increase in following — in fact most news channels added double the number of users they add in a day. An interesting case is that of Kangana Ranaut (@KanganaTeam) which was losing followers in the days preceding Feb 2 but added many followers to what may be attributed to the series of tweets from the account in response to the Rihanna tweet at first, and a subsequent online spat with actor Tapasee Pannu.
\begin{table*}[!htb]
\small
\begin{tabular}{lccc}
\hline
\textbf{Username} & \begin{tabular}[c]{@{}c@{}c@{}}\textbf{increase in} \\\textbf{followers from} \\ \textbf{31st Jan to 2nd Feb}\end{tabular} & \begin{tabular}[c]{@{}c@{}c@{}}\textbf{increase in}\\ \textbf{followers from}\\ \textbf{2nd to 4th Feb}\end{tabular} & \begin{tabular}[c]{@{}c@{}c@{}}\textbf{increase over} \\ \textbf{corresponding} \\ \textbf{
two-day period}\end{tabular} \\ \hline
rihanna & 95293 & 451794 & 356501(375\%) \\ \hline
miakhalifa & 3797 & 63817 & 60020(1581\%) \\ \hline
akshaykumar & 16047 & 48244 & 32197(201\%) \\ \hline
imVkohli & 25079 & 55185 & 30106(121\%) \\ \hline
AmitShah & 17576 & 45272 & 27696(158\%) \\ \hline
KanganaTeam & -1370 & 25417 & 26787(NA) \\ \hline
ImRo45 & 9362 & 32076 & 22714(243\%) \\ \hline
myogiadityanath & 13968 & 34703 & 20735(149\%) \\ \hline
sachin\_rt & 8999 & 28087 & 19088(213\%) \\ \hline
hrw & 4560 & 21654 & 17094(375\%) \\ \hline
sudhirchaudhary & 3038 & 17880 & 14842(489\%) \\ \hline
ImRaina & 8173 & 22679 & 14506(178\%) \\ \hline
ajinkyarahane88 & 9219 & 22472 & 13253(144\%) \\ \hline
AnupamPKher & 3547 & 16268 & 12721(359\%) \\ \hline
hardikpandya7 & 8207 & 20382 & 12175(149\%) \\ \hline
MEAIndia & 266 & 11525 & 11259(4233\%) \\ \hline
ajaydevgn & 3379 & 13963 & 10584(314\%) \\ \hline
dhruv\_rathee & 4902 & 14997 & 10095(206\%) \\ \hline
\end{tabular}
\caption{\label{tab:Table 2}\textbf{Increase in Following Among Key Figures Who Tweeted About the Issue} (Note: Column 4 depicts percent increase in following relative to increase in following over earlier days)}
\end{table*}

In a nutshell what we see is a confirmation of much past research that controversial behavior helps rather than hinders the social media appeal of those involved. Perhaps over and above all, it confirms that angry exchanges benefit the platform above all, since following went up across the board.

\section{Misinformation}
\begin{figure}[!htb]
  \centering
  \includegraphics[width=\linewidth]{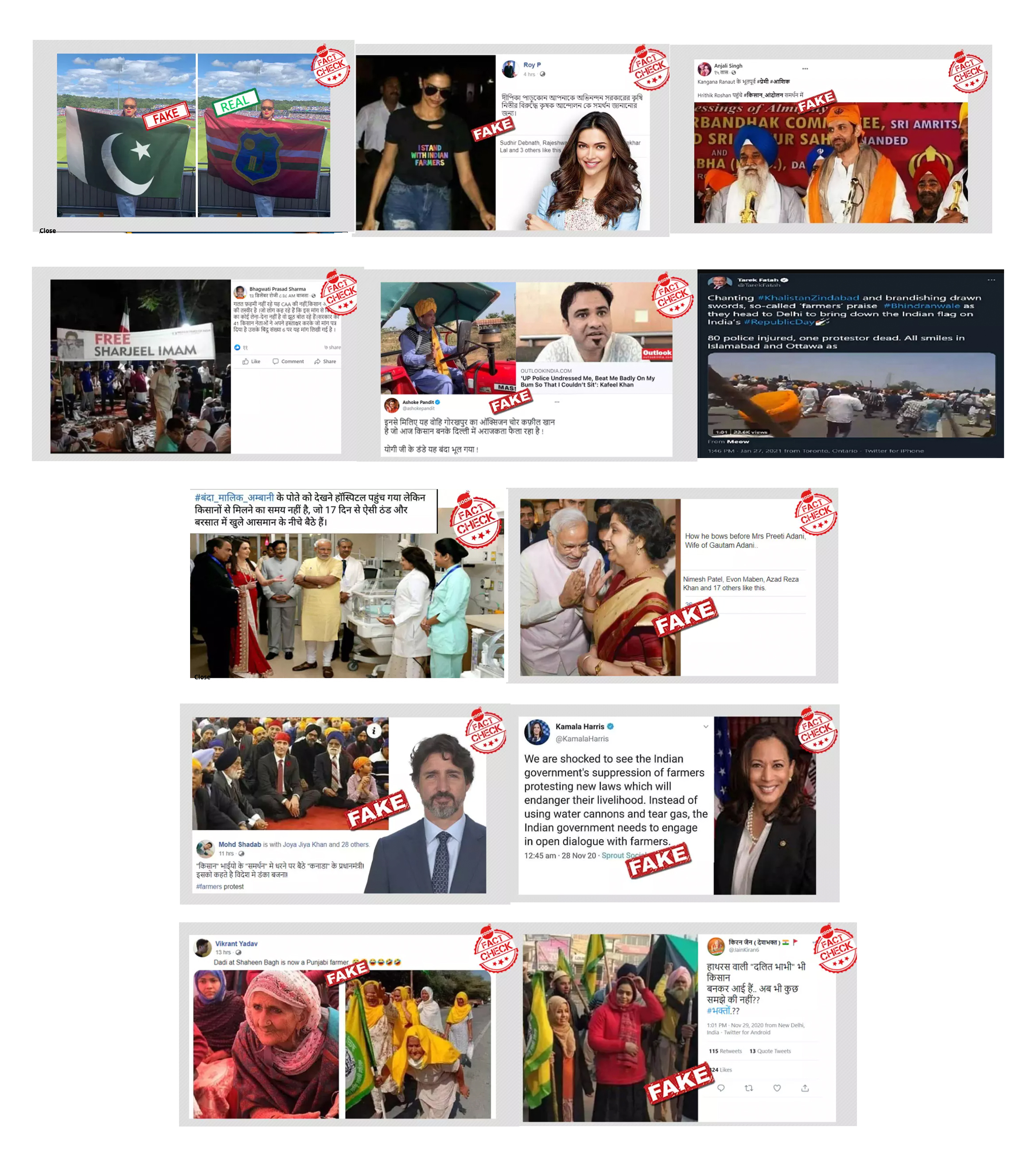}
  \caption{\textbf{Various Actors Associated in Debunked Stories on Farmers’ Protest} (Source: Boomlive)}
\end{figure}
Finally, the misinformation related to the farmers’ protests is a necessary part of understanding what has driven some of the polarization that we see in this case. The misinformation on this issue goes back several months and has been used to throw shade on both the government and the farmers.

The first set of debunked stories on the farmers’ protests in September 2020, started by falsely depicting a massive number of farmers on the ground to protest the Agriculture-related bills passed by the central government. These stories saw images from older protests such as the Anti-CAA/NRC protests, being recontextualized with a newer narrative. While fake stories are mostly used to show someone in a bad light, here they were used in a positive light to present a false perception of strength and power. These stories were quickly seen evolving into showing false depictions of police brutality in the otherwise peaceful protests. They included graphic content of deceased farmers at the protest site and women beaten to ground, which is a commonly used tactic to evoke emotional responses from the audience.

The second and most frequent target of the trolls was prime minister Narendra Modi himself. He was portrayed in relation to the Adanis and Ambanis, where he is depicted bowing to Priti Adani in seeming obeisance or visiting a grandchild of the Ambani family. Another set of debunked stories related to Narendra Modi falsely showed the people of the Sikh community disrespecting him. Trolls often use well know actors to promote or dissuade the current topic of interest. For instance, celebrities like Hrithik Roshan and Deepika Padukone were falsely shown to be supporting the farmers’ protests.  They have become frequent targets of the trolls – after they were vocal about their anti-CAA stance – to depict an anti-government agenda. Similarly, Rihanna was targeted using a photoshopped image where she is seen holding Pakistan’s flag.

Other widely recognized public persons such as Bilkis Dadi, a senior woman who took part in the Anti-CAA protests, and `Hathras Bhabhi' a relative of the victim's family from the Hathras rape case, falsely emerged as supporters of the farmer’s protest. This was ostensibly done to present the notion that individuals involved in one or another protest against the government double up in others as well, thereby delegitimizing their causes as being driven by professional protestors. International politicians such as Kamala Harris and Justin Trudeau were also bought in as supporters of the farmers using photoshopped images. Various actors mentioned above were used to promote a false narrative that these protests are part of an international propaganda and conspiracy.

The narratives further turned insidious when the farmers were falsely depicted to be pro-Khalistan members. Various photographs were seen to be photoshopped and recontextualized to depict farmers as anti-India sloganeers. These stories included photoshopped banners of the Khalistan flag and older footage from pro-Khalistan rallies wrongly associated with current protests. Twitter accounts that have on multiple occasions been party to spreading misinformation, such as that of journalist Tarek Fateh, chimed in branding farmers as pro-Khalistan. The major story in this segment occurred on republic day, the 26th of January, were the farmers entered the red fort. It was wrongly being spread that the farmers had replaced the Indian flag with the flag of Khalistan. The Indian flag is repeatedly used as a tactic by trolls to paint an anti-national narrative. In various stories, the farmers have been falsely depicted to be disrespecting the flag by burning and stamping on it.

The trends of conspiracy against Muslim community continued to emerge in relation to the protests. A communal spin was propagated against the farmers where Muslims were shown to be disguised as Sikhs and people from the Sikh community were shown praying namaaz. Muslim women protestors in burqas were also shown amidst the farmers. Other famous Muslim men such as activist Sharjeel Imam and pediatrician, Kafeel Khan, both political prisoners, falsely associated with the farmers. In these references, the idea put forth is that the farmers are being supported by those already been branded as `anti-nationals'.

Finally, a new tactic of using fake twitter profiles was also noticed. This tactic uses fake profile pictures, fake screenshots of tweets or newly created fake accounts itself to show an individual’s conspicuous association or ill intention. Here, fake twitter handles of Rakesh Tikait – leader of a farmers' union- were seen to be created and certain other handles were seen to misuse pictures of Pakistani women to frame false narratives.

\section{Conclusion}
\begin{figure}[!htb]
  \centering
  \includegraphics[width=\linewidth]{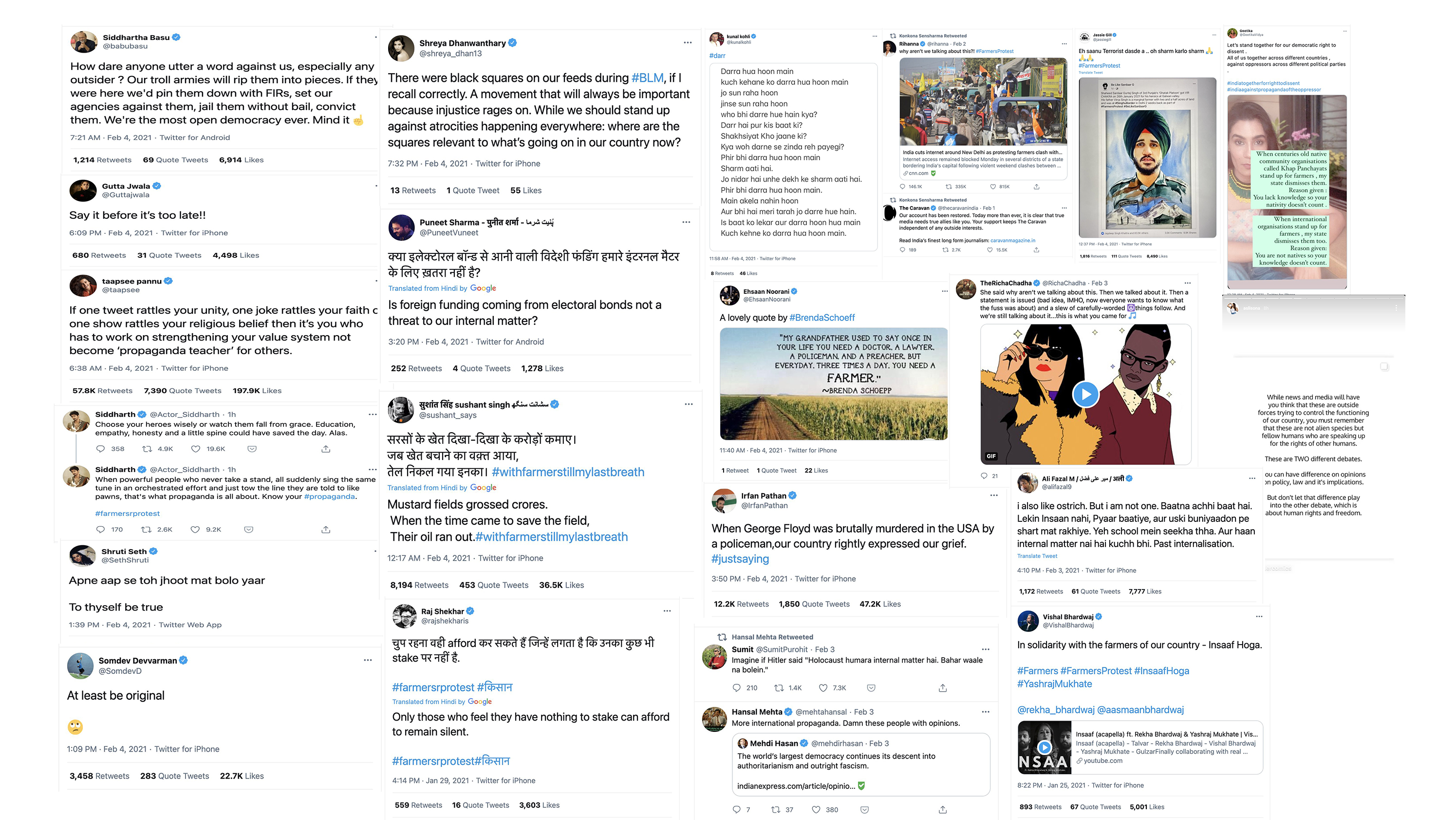}
  \caption{\textbf{A Collage of Celebrities Who Came Out in Support of Farmers' Protest}}
\end{figure}
While some of the patterns seen here may present a picture that the overwhelming majority of major public figures align themselves with the official line, or avoid engaging entirely (for instance cricketers MS Dhoni and Rahul Dravid both trended for not being active on twitter), the incident has also thrown open one of the most significant cases of open confrontation between a subset of celebrities – mainly entertainers and sportspersons who have come out against the government. This was largely unseen in the past or restricted to a small number of outspoken individuals.

While cricketers, in particular, have been on the side of the government through this, the event also marked a rare case in which one of the sport’s most beloved legends, Sachin Tendulkar, winner of the highest civilian honour in the country, came under open derision on social media. Indeed, in a rare departure, Irfan Pathan, one of India’s top 10 career wicket takers in One Day Internationals, came out with a tweet that said “When George Floyd was brutally murdered in the USA by a policeman, our country rightly expressed our grief. \#justsaying”

\section{Acknowledgements}
The authors would like to thank Tattle\cite{Tattle} for their archive of debunked misinformation. The work presented here is solely that of the authors and does not represent the views of their organizations.


\bibliographystyle{ACM-Reference-Format}
\bibliography{sample-base}










\end{document}